\documentclass[conference]{IEEEtran}
\IEEEoverridecommandlockouts
\PassOptionsToPackage{}{hyperref}
\PassOptionsToPackage{obeyspaces}{url}

\usepackage{float, enumerate, xcolor, algorithm, algpseudocode, tabto, balance, amsmath, graphicx, hyperref, multirow, dblfloatfix, listings, eurosym, lstautogobble, multirow, tikz, comment} 
\usepackage[table]{colortbl}
\definecolor{cellFillingColor}{rgb}{0.8, 0.8, 0.8}
\usepackage{amsfonts}
\usepackage{amssymb}
\usepackage{booktabs}
\usepackage{url}
\usepackage{amssymb}
\usepackage[numbers,sort&compress]{natbib}
\usepackage{subfigure}
\usetikzlibrary{arrows,backgrounds,positioning}
\usepackage{chngcntr}
\usepackage[overload]{empheq}
\usepackage{soul}
\usepackage[multiple]{footmisc} 

\usepackage{array}
\newcolumntype{L}{>{\centering\arraybackslash}m{.5\columnwidth}|}

\usepackage{url}

\usepackage{pifont}
\newcommand{\cmark}{\ding{51}} %BOLD CHECK
\newcommand{\xmark}{\ding{55}} %BOLD CROSS
\usetikzlibrary{arrows,automata}

\newcommand*\circled[1]{\tikz[baseline=(char.base)]{\node[shape=circle,
		draw,
		inner sep=0.0pt,
		text width=3mm,
		align=center, 
		] (char) {#1};}}
\newcommand*\rot{\rotatebox{90}}
\graphicspath{ {./images/} }
\def\BibTeX{{\rm B\kern-.05em{\sc i\kern-.025em b}\kern-.08em
		T\kern-.1667em\lower.7ex\hbox{E}\kern-.125emX}}

\begin{document}
	\title{De-authentication using Ambient Light Sensor}
	\author{
		\IEEEauthorblockN{Ankit Gangwal}
		\IEEEauthorblockA{
			\textit{International Institute of Information}\\\textit{Technology Hyderabad, India}\\
			gangwal@iiit.ac.in}
		\and
		\IEEEauthorblockN{Aashish Paliwal}
		\IEEEauthorblockA{
			\textit{International Institute of Information}\\\textit{Technology Hyderabad, India}\\
			aashish.paliwal@research.iiit.ac.in}
		\and
		\IEEEauthorblockN{Mauro Conti}
		\IEEEauthorblockA{
			\textit{University of Padua, Italy}\\
			mauro.conti@unipd.it}
		\thanks{This paper has been accepted for publication in IEEE ACCESS.}
		\thanks{DOI: 10.1109/ACCESS.2024.3367607}
	}
	\maketitle
	
	\begin{abstract}
		While user authentication happens before initiating or resuming a login session, de-authentication detects the absence of a pre-viously-authenticated user to revoke her currently active login session. The absence of proper de-authentication can lead to well-known \textit{lunchtime} attacks, where a nearby adversary takes over a carelessly departed user's running login session. The existing solutions for automatic de-authentica-tion have distinct practical limitations, e.g., extraordinary deployment requirements or high initial cost of external equipment.
		\par
		In this paper, we propose ``DE-authentication using Ambient Light sensor''~(DEAL), a novel, inexpensive, fast, and user-friendly de-authentica-tion approach. DEAL utilizes the built-in ambient light sensor of a modern computer to determine if the user is leaving her work-desk. DEAL, by design, is resilient to natural shifts in lighting conditions and can be configured to handle abrupt changes in ambient illumination (e.g., due to toggling of room lights). We collected data samples from 4800~sessions with 120~volunteers in 4~typical workplace settings and conducted a series of experiments to evaluate the quality of our proposed approach thoroughly. Our results show that DEAL can de-authenticate a departing user within 4~seconds with a hit rate of~89.15\% and a fall-out of 7.35\%. Finally, bypassing DEAL to launch a \textit{lunchtime} attack is practically infeasible as it requires the attacker to either take the user's position within a few seconds or manipulate the sensor readings sophisticatedly in real-time.
	\end{abstract}


	\section{Introduction}
	\label{section:introduction}
	Computer users in different establishments~(e.g., universities, workplaces) often share workspace. These users either work on shared computers~(e.g., in a library) or have a dedicated computer\footnote{We use the term `computer' to equally represent a desktop and a laptop.}~(e.g., in an office). In either case, user authentication is critical to prevent any unauthorized access. Generally, the user authenticates via the secret PIN, password, or recently emerging biometrics-based techniques. However, such authentication typically happens only once while initiating the login session. After successful authentication, the user spends time to continuously use the computer and its services. If the user wants to leave her computer for whatever reason during this period, her currently active session must be locked/logged out; especially in shared workspace settings. Failure to do so can lead to \textit{lunchtime} attacks~\cite{eberz2015preventing, kaczmarek2018assentication}, where an adversary~(typically an insider) gains access to the user's running session and engages in potentially undesirable activities.
	\par
	To prevent such unauthorized access, either the user must terminate the running session by explicitly locking/logging out, or the system must automatically revoke the previously-authenticated session, i.e., de-authenticate the user. Oftentimes, the users are apathetic or lazy~(especially when taking short breaks) and avoid terminating the session because logging in again can be annoying. On another side, de-authenticating the user frequently with too-short inactivity timeouts can aggravate the user while choosing a too-long inactivity timeout leaves room for \textit{lunchtime} attacks~\cite{sinclair2008preventative}.	
	\par
	Researchers from both academia and industry have put immense efforts into making the authentication techniques more robust, accurate, efficient, and convenient to use~\cite{wang2021attacks}. For instance, biometric-based authentication techniques impose less cognitive load on the users compared to the password-based approach. Nonetheless, password-based authentication still remains the most commonly used approach; mainly because it is intuitive and does not require any special hardware. But passwords have their demerits. First, recent technological advances are making passwords even more susceptible to cracking and potentially obsolete for use in the near future~\cite{pasquini2021improving}. Second, passwords have no role in automatic user de-authentication, which means a separate mechanism is required. To this end, researchers have proposed different user de-authentication and continuous-authentication mechanisms. The state-of-the-art solutions~(cf. Section~\ref{section:relatedWork}) require external equipments~\cite{kaczmarek2018assentication, eberz2015preventing, rasmussen2014authentication, conti2017fadewich, conti2020auth, mare2014zebra, corner2002zero, ecg}, are relatively expensive~\cite{eberz2015preventing, mare2014zebra, ecg, kaczmarek2018assentication}, need physical customization or specific installation~\cite{eberz2015preventing, kaczmarek2018assentication, conti2017fadewich, conti2020auth, blufade}, are complex to deploy~\cite{kaczmarek2018assentication, conti2017fadewich, rasmussen2014authentication}, involve regular maintenance~\cite{kaczmarek2018assentication, conti2020auth, corner2002zero, mare2014zebra}, or sometimes cause inconvenience to the user~\cite{rasmussen2014authentication, corner2002zero, mare2014zebra, ecg}. Such limitations hinder a broader adoption of the existing solutions. Therefore, a solution is needed that can address all of these issues while handling the automatic user de-authentication process efficiently.
	\par
	On the other side, consumer devices~(e.g., phones, tablets, computers) are becoming sensor-rich to provide different useful functionalities. Ambient Light Sensor~(ALS) is one such sensor. ALS has been pervasively found on phones and tablets. Nonetheless, ALS has recently started to become common on consumer-grade laptops and desktops; primarily to comfort users' eyes by adapting the brightness and/or color tone of the screen in response to changing lighting conditions. ALS is generally mounted on a computer's display screen~(e.g., as shown in \figurename~\ref{fig:concept}). A generic ALS is both fast and efficient in capturing changes in lighting conditions.
	\par
	In this paper, we propose ``DE-authentication using Ambient Light sensor''~(DEAL), a novel de-authentication technique that utilizes the built-in ALS of a computer to decide whether the user is leaving her work-desk. In particular, DEAL takes advantage of the fact that a user normally sits/stands closer~(suggested between 16 to 30~inches~\cite{eyes}) to the computer while working. Thus, the user can affect the illumination perceived by the computer's ALS when she moves away. In the simplest case, the user directly blocks the Line-of-Sight~(LoS) path between ALS and the light source. Nonetheless, the ambient lighting conditions around the computer's ALS can also be influenced due to partial blocking of its LoS, shadowing it, or even reflection of light towards it from the departing user's body~(cf. Section~\ref{section:proposedMethod}). We design DEAL to analyze the changes in lighting conditions via ALS readings to decide whether the user is departing from her work-desk. DEAL intrinsically addresses the above-mentioned issues of the state-of-the-art works by its design, i.e.,~(1)~no external equipment is required as ALS is built-in a modern computer, (2)~ALS is low-cost that too is already included in the computer's cost, (3)~no physical installation of hardware is needed, (4)~it is simple to deploy its software, (5)~no periodic maintenance is required as an ALS is generally long-lasting and is powered directly by the computer, and (6)~more importantly, it is user-friendly as the user is not required to carry or wear any apparatus.
	\par
	\textit{Contribution:} The contributions of our work are as follows:
	\begin{enumerate}
		\item We propose DEAL, a novel, unobtrusive, fast, and inexpensive de-authentica-tion approach that is suitable for modern computers equipped with a built-in ALS.
		\item We thoroughly evaluate the performance of our proposed approach using data samples collected from 4800~sessions with 120~volunteers in 4~typical workplace settings. DEAL can attain an overall hit rate of~89.15\% and a fall-out of 7.35\% to de-authenticate the user within 4~seconds.
		\item Finally, we compare DEAL with the state-of-the-art de-auth-entication approaches and delineate their respective advantages and limitations. We argue that the said performance of DEAL comes without any extraordinary requirements, customization, or expensive equipment, which makes it suitable for practical adoption.
	\end{enumerate}
	\par
	\textit{Organization:} The remainder of this paper is organized as follows. Section~\ref{section:relatedWork} presents a comparative summary of the related works. We elucidate our system and adversary models in Section~\ref{section:systemAdversaryModel}. We explain our proposed approach in Section~\ref{section:proposedMethod} and present its evaluation in Section~\ref{section:evaluation}. Section~\ref{section:discussion} elaborates on the salient features and potential limitations of our work. Finally, Section~\ref{section:conclusionFutureWork} concludes the paper.

	\section{Related works}
	\label{section:relatedWork}
	Researchers from both academia and industry have put extensive efforts over the decades to develop effective user authentication techniques. To verify a user's identity, a typical authentication procedure utilize:~(1)~user's knowledge~(e.g., password, pin)~\cite{pasquini2021improving}, (2)~user's possession~(e.g., token, keycard)~\cite{berenjestanaki2019exploitation}, (3)~user's physical attributes~(e.g., biometrics)~\cite{banerjee2012biometric}, (4)~user's behavior~(e.g., gestures, typing patterns, eye movements)~\cite{kinnunen2010towards}, or (5)~a combination of these to enable two-factor authentication~\cite{bonneau2012science, cnssi2015}.
	\par
	On another side, the need of user de-authentication arises after successful authentication of a user by the system. It is worth mentioning that a user's de-authentication by the system is independent of the authentication step. Therefore, the procedures for user de-authentication are distinct. One of the commonly used mechanisms for user de-authentication is the inactivity time-out approach. However, such an approach is ineffective because:~(1)~determining the optimal length of a static timeout interval is not straightforward, and (2)~checking the user's presence/absence in front of the system is beyond its scope~\cite{sinclair2008preventative}. Given the significance of user de-authentication to prevent \textit{lunchtime} attacks, different mechanisms have been proposed that aim at continuously establishing the user's presence/absence near the system. 
	\par
	Kaczmarek~et~al.~\cite{kaczmarek2018assentication} propose \textit{Assentication} to profile user's sitting posture. In particular, \textit{Assentication} installs 16 pressure sensors in an office chair to capture a hybrid biometric trait by combining user's behavioral and physiological characteristics. Though \textit{Assentication} has low false positive and false negative rates, it has two key limitations. Firstly, it has low permanence, i.e., the hybrid biometric trait that it captures naturally changes over time for a given user. Secondly, the cost involved is not trivial, i.e., about \$150 per chair. While eye movement tracking has been previously employed to authenticate users~\cite{kinnunen2010towards}, Eberz~et~al.~\cite{eberz2015preventing} use gaze tracking to prevent \textit{lunchtime} attacks. Their system continuously tracks the user's eye movements with high accuracy. Since gaze tracking requires its user to keep their sight in a particular direction, any head moment taking the sight away can generate false positives. Moreover, the cost of eye-tracking equipment hinders its large-scale adoption. Rasmussen~et~al.~\cite{rasmussen2014authentication} propose a new biometric based on the human body's response to an electric pulse signal. Their approach involves applying a low-voltage pulse signal to user's one palm and measuring the body's response in the user's other palm. Apart from the cost of specialized hardware, engaging both hands of the users with pulse-response hardware restricts its general acceptability. Similarly, authors in the work~\cite{ecg} use ECGs to build continuous authentication systems that require end users to wear specialized hardware.
	\par
	FADEWICH~\cite{conti2017fadewich} measures the attenuation of wireless signals due to the human body for estimating the location of a user in a room, and the user is de-authenticated based on the user's estimated position. Their system uses 9 sensors in a fixed office setup to achieve very high accuracy. The major drawback of their approach is that the structure and setup of the office heavily affect the placement of sensors. Thus, each office requires customized positioning of sensors. Moreover, the presence and movements of other persons induce false positives. Keystroke dynamics technique~\cite{joyce1990identity} profiles a user's typing style. It is a simpler mechanism for continuous authentication, which is easily deployable and does not need specialized hardware. However, researchers~\cite{tey2013can} have shown that a brief training is sufficient to imitate typing pattern of the target users, even when their typing patterns are only partially known. DEB~\cite{conti2020auth} instruments an office chair with two Bluetooth low-energy beacons. An application running on the target system monitors the signal strength of the received Bluetooth beacons. A human body present in the line of sight of a beacon affects the strength of the received signal, which is interpreted to keep the user logged into the system. Apart from interference due to nearby beacons, the lifespan and appropriate installation of Bluetooth beacons are the key concerns here. BLUFADE~\cite{blufade} employs deep learning algorithms to continuously detect the user's face in a webcam feed. However, using a camera feed for de-authentication carries apparent privacy concerns~\cite{privacyReportHP}. Thus, the authors propose to obfuscate the webcam with a physical blurring layer (e.g., anti-reflective obfuscating film) and use blurred images for face detection. Such an approach hampers the normal usage of the webcam. More importantly, it does not address the possibility of reconstructing the user's facial traits from blurry images.
	\par
	ZIA~\cite{corner2002zero} proposes monitoring the proximity of the user via a physical token borne by the user. Such a token periodically exchange information with the target system over a secure channel, and in the absence of such communication the user is de-authenticated by the system. Similarly, ZEBRA~\cite{mare2014zebra} uses a wrist bracelet fitted with a gyroscope, an accelerometer, and a radio. When the user interacts with the system, the bracelet captures and shares the wrist movements with an application running on the system. The application correlates the wrist movements with strokes on the keyboard to establish the user's presence. The key limitation of these approaches is that the user is required to always bear the token/bracelet. Furthermore, the tokens/bracelets also require periodic recharging or replacement of batteries. Relevant to our work, researchers have used ALS for user authentication~\cite{yoon2015exploiting} and tracking a user's activities~\cite{holmes2016luxleak, ionescu2016exploiting, spreitzer2014pin, shang2020lightdefender}.

	\section{System and adversary models}
	\label{section:systemAdversaryModel}	
		In this section, we describe the system and adversary models we consider in our work. Section~\ref{section:systemModel} presents the deployment scenario of the proposed de-authentication mechanism, and Section~\ref{section:adversaryModel} elucidates the potential threat maneuvers of an adversary.

		\subsection{System model}
		\label{section:systemModel}
		DEAL is designed for computers that come with a built-in ALS. The ALS data feed is processed in real-time by a simple application running in the background on the target computer. Since the primary goal of any de-authentication mechanism is to prevent \textit{lunchtime} attacks that are prevalent at workplaces~\cite{eberz2015preventing, conti2020auth, kaczmarek2018assentication}, our proposed system is expected to be used in common workplace setups. DEAL is absolutely unobtrusive. The user arrives at her work-desk, settles in her chair, logs into her computer~(a desktop or docked laptop) via a preset authentication mechanism, uses the computer, and finally gets up to leave her desk. While the user prepares to depart from her desk, the system should automatically lock her out to prevent any unauthorized access. DEAL uses the light-intensity data feed from ALS to de-authenticate a departing user in real-time. 
		\par
		Contrary to state-of-the-art de-/continuous-authentication mechanisms~\cite{eberz2015preventing, rasmussen2014authentication, joyce1990identity, mare2014zebra}, DEAL does not need the user to interact continuously with the system. In fact, there can be situations when the user is present at the work-desk, but not interacting with the system. For instance, the user may be using a smartphone, reading a document, or simply watching a photo on the system. In such scenarios, de-authenticating the user due to her inactivity is undesirable and can be annoying.
		
		\subsection{Adversary model}
		\label{section:adversaryModel}
		We assume that the adversary has physical access to the user's office and, consequently, to her computer. An office colleague, a visitor in the office, a business customer, or a housekeeping person are some representative examples of such an adversary that may be interested in getting access to her computer. Since the adversary does not know the login credentials required for logging in to the user's computer, the adversary's goal is to gain access to the user's running/authenticated session.
		\par
		An adversary can try the following to bypass DEAL:~(1)~take the user's position~(and control the computer) before DEAL can de-authenticate the user, or (2)~manipulate light intensity perceived by her computer's ALS in such a way that DEAL does not de-authenticate the departing user at all. The former approach represents the typical \textit{lunchtime} attack strategy. It is straightforward, yet effective if DEAL takes too long to de-authenticate. So, DEAL should operate fast enough to render such an attempt ineffective. The latter may involve using sophisticated tools. For instance, the adversary may use a custom beam of light to compensate for the ALS readings affected due to the departing user. Such a maneuver requires the adversary to know the exact light intensity levels observed by the target ALS when the user is departing, which may be possible by:~(1)~installing an ALS near the target's ALS~(ineffective; as it will visible to the user), (2)~compromising the target machine to get such information~(beyond the scope of the \textit{lunchtime} attack), or (3)~physically approaching the desk to measure/compensate readings~(essentially the same as the first approach; a fast operating DEAL will handle it). 
		\par
		On another side, a different type of adversary can focus on triggering false de-authentications, e.g., \textbf{by turning the lights on or off in the room.} Although such an action can annoy the user by incorrectly de-authenticating her, the adversary does not get access to the user's computer. Nonetheless, toggling room lights is a part of routine office activities. Such sudden changes in lighting conditions significantly affect the ALS readings and induce large outliers. Therefore, we can easily identify and adapt to new lighting conditions if such large outliers in the ALS readings are consistently present.
		
	\section{Proposed Method}
	\label{section:proposedMethod}
\begin{figure}[H]
		\centering
		\includegraphics[trim =0mm 4mm 0mm 0mm, clip, width=0.6\columnwidth]{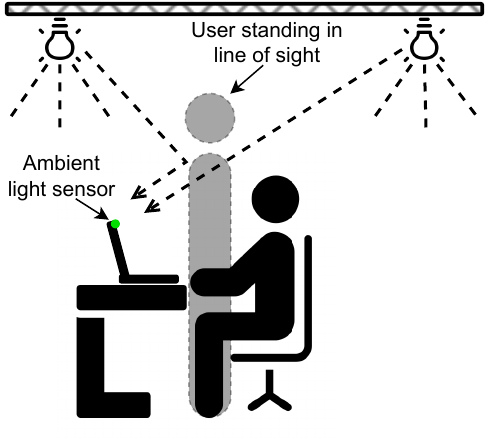}
		\caption{A representative depiction of affecting illumination perceived by ALS.}
		\label{fig:concept}
	\end{figure}

	We now present the conceptual and intrinsic details of DEAL. The fundamental task of a meaningful de-authentication technique is to determine the user's presence in front of the computer. To this end, DEAL utilizes data feed from the computer's ALS. The illumination perceived by ALS can be affected due to the user's movements. As a representative example, \figurename~\ref{fig:concept} demonstrates that a user's movement of getting up/down from her chair can directly affect the ambient lighting conditions around the computer's ALS. Naturally, the scale and duration of such an impact depends on a variety of factors, e.g., how much/for how long the user has intersected the LoS path between ALS and the light source. We would like to highlight that though the light sources are typically roof-mounted~(or, mounted high on the wall) in workplaces, the light source may not be in the direct LoS of ALS~(cf. \figurename~\ref{fig:concept}). However, a user's movements can still affect the lighting conditions around ALS. In particular, due to partial\footnote{In full blocking, the user totally obstructs the illumination received by ALS. The simplest example would be to cover ALS by hands. In partial blocking, the user partially hinders the light coming from a source. For instance, when a user intercepts ALS's LoS partially. The shadow of the user may or may not be falling around ALS in partial blocking. We call the former scenario shadowing, and the reflection of light is a natural phenomenon.} blocking, shadowing, or even reflection of light from the departing user's body.	By measuring the changes in ambient lighting conditions through ALS readings, DEAL determines whether the user is departing from her work-desk, and subsequently de-authenticates her when required.
	\par
	We make the following two reasonable assumptions in the implementation of DEAL:~(1)~the user will continue to work in the same position~(standing or sitting) as she was in while initializing the current login session, and (2)~if the user was sitting, she will get up before leaving. It is worth mentioning that if the user is standing while working at her work-desk, she will likely be blocking ALS' LoS. Such a case is simpler to handle for DEAL. For the sake of brevity, the rest of the paper proceeds with the scenario in which the user is sitting while using her computer.
	\par	
	The data feed from ALS can be modeled as a univariate time series of the observed light intensity. Thus, DEAL adopts an amended sliding window average-based approach for monitoring changes in lighting conditions. In particular, each reading~($R$, in lux) from ALS is compared against the average of running \textit{window} as described in Eq.~\ref{eq:slidingWindow}:
	\begin{equation}
		\label{eq:slidingWindow}
		|\mu(window)-R| > \mu(window)*\Delta/100,
	\end{equation}
	where $\Delta$~(a natural number) is a predefined threshold. Such sliding window average-based methods are typically designed to identify an outlier outside of the current trend in a time series. However, a single outlier may not be sufficient in our case to distinguish the user's movements correctly. Because an ALS can provide several readings - according to its operating frequency~($f$, in Hz) - within a fraction of time. Moreover, different user activities can last for a different amount of time, e.g., the act of getting up and moving away from the computer can take up to a few seconds. So, it is intuitive to say that if a user's movement intercepts ALS's LoS for a longer period of time, then it will affect more ALS readings. We design DEAL to incorporate the duration of impact on ALS to distinguish user movements. To this end, we define a parameter $\eta$~(in seconds). While $\Delta$ defines the minimum distance between an outlier and the average of running \textit{window}~(cf. Eq.~\ref{eq:slidingWindow}), $\eta$ specifies the minimum duration of time during which each consecutive\footnote{Our current implementation requires each consecutive $R$ in $\eta$ duration of time to be an outlier. We are aware that such a design decision can result in false negatives even when one of the values is not an outlier. However, such a stricter control helps us evaluate the minimum performance of our system. We can certainly optimize such checks to improve the system. Currently, $\eta$ works with a parameter $\ell$ to provide some relaxation to the system.} $R$ should be an outlier for recognizing the user to be departing and subsequently de-authenticating her. 
	\par
	Our system has two more parameters, i.e., $\omega$~(in seconds) and $\ell$~(in seconds). $\omega$ defines the size of the sliding window. $\ell$ is a tuning parameter that defines the maximum duration of time from the occurrence of the first outlier in a wave of outliers, during which the required consecutive outliers should occur for user de-authentication. From the virtues of their respective definitions, $\eta \leq \ell$. The system will not work if $\eta > \ell$, because it is impossible to have $\eta$~(say 5 seconds) of consecutive outliers within a shorter $\ell$~(say 2 seconds). To simplify, $\ell$ separates waves of outliers. A larger value of $\ell$ enables us to process more values of $R$ to satisfy constraints on $\eta$. However, a larger $\ell$ will cause a delay in resetting and recovering from a~(short) wave of outliers. On another side, a larger value of $\eta$ prevents false alarms due to subtle user movements. Algorithm~\ref{lst:algo1} exhibits the pseudocode for the core logic of DEAL.
	\begin{algorithm}[!t]
		\caption{A simplified pseudocode for DEAL's core logic.}
		\label{lst:algo1}
		\hspace*{\algorithmicindent} \textbf{Input:} $\Delta$, $\omega$, $\eta$, $\ell$
		\begin{algorithmic}[1]
			\State $f := $ \textit{FreqALS()} \Comment{Sample ALS's operting frequency}
			\State $\omega^\prime := int(\omega * f)$ \Comment{Align $\omega$ to ALS via $f$}
			\State $\eta^\prime := int(\eta * f)$ \Comment{Align $\eta$ to ALS via $f$}
			\State \textit{window} $ := $ List with recent $\omega^\prime$ ALS readings
			\State $temp_1 := 0$ \Comment{Tracks number of consecutive outliers}
			\State $temp_2 := 0$ \Comment{Stores time of first outlier in current wave}
			
			\While{\textit{true}}
			\State $R := $  \textit{readALS()}
			\If{\textit{(abs($\mu$(window)-R)} $>$ \textit{$\mu$(window)}*$\Delta$/100)}
			\State $temp_1 := temp_1+1$
			\If{$temp_2 =0$}
			\State $temp_2 := getTime()$
			\EndIf
			\Else
			\State $temp_1 := 0$
			\State \textit{window.append(R)} \Comment{Append \textit{R} to \textit{window}}
			\State \textit{window.pop(0)} \Comment{Pop tail from \textit{window}}
			\EndIf
			\If{$(temp_1 \geq \eta^\prime$ \&\& $(getTime()-temp_2) \leq \ell$)}
			\State \textit{De-authenticate()}
			\EndIf
			\If{$(temp_2 \neq 0$ \&\& $(getTime()-temp_2) > \ell$)}
			\State \textit{reSet(window,} $temp_1$\textit{,} $temp_2$\textit{)} \Comment{Go to line 4}
			\EndIf
			\EndWhile
		\end{algorithmic}
	\end{algorithm}
	\par
	Since each ALS can operate at a different $f$, we begin with aligning $\omega$ and $\eta$ to a given ALS via its $f$~(lines~1-3). We next initialize the \textit{window} and temporary variables~(lines~4-6). We compare each reading from ALS with the mean of \textit{window}~(lines~7-9). If an outlier is found, a counter is incremented~(line~10) while the time is recorded for the first outlier~(lines~11-12). If $R$ is not an outlier, the counter for outliers is reset~(line~15), and $R$ is adjusted in \textit{window}~(lines~16-17). \textbf{It is noteworthy that the running \textit{window} average directly handles the natural shifts in lighting conditions.} The user is de-authenticated if the required number of outliers~($\eta^\prime$) are found within $\ell$~seconds~(lines~19-20). If the time elapsed since the first outlier in the current wave was seen exceeds $\ell$, we reset \textit{window} and temporary variables~(lines~22-23).

	\section{Evaluation}
	\label{section:evaluation}
	We describe our evaluation setup in Section~\ref{section:evalSetup} and data collection method in Section~\ref{section:dataCollection}. We discuss our experimental results in Section~\ref{section:results}.
	
	\subsection{Evaluation setup}
	\label{section:evalSetup}
\begin{figure}[!htbp]
		\centering
		\includegraphics[trim =0mm 2.5mm 0mm 0mm, clip, width=0.9\columnwidth]{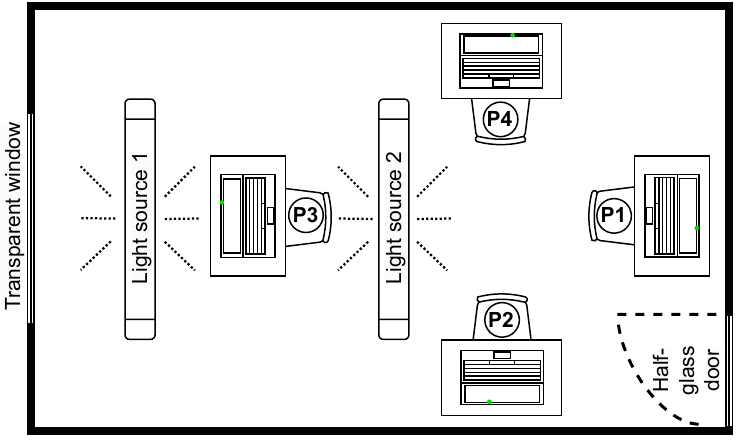}
		\caption{The top view of our office setup.}
		\label{fig:roomOutline}
	\end{figure}

	We evaluate DEAL in a typical office setup. To this end, we created an office space illuminated with both natural and artificial lights. As shown in \figurename~\ref{fig:roomOutline}, our office setup has two ceiling-mounted white light sources that we keep on and a standard transparent window that allows natural light to come in. Though the half-glass door was kept closed during the experiments, its transparent glass portion in the upper half remained unobstructed.
	\par
	To emulate typical work-desk positions with respect to the lighting conditions, we set up four work-desks at different locations in the room~(cf. \figurename~\ref{fig:roomOutline}). In particular, position~\textit{P1} emulates a position with lower lighting since it is far from the light sources. Moreover, a user working in position~\textit{P1} may further block the illumination perceived by the computer's ALS. Being closer to light source~2, positions~\textit{P2} and \textit{P4} represent normally illuminated positions. Lastly, position~\textit{P3} has copious lighting. In our experiments, we used a Lenovo ThinkPad Yoga 370 laptop. It comes with a built-in ALS. We modified \textit{iio-sensor-proxy}~\cite{iioSensorProxy} to capture readings from ALS. It is important to highlight that we periodically checked the health of our computer's ALS using a phone-based ALS to avoid any bias or error in our ALS readings.
	
	\subsection{Data collection}
	\label{section:dataCollection}	
\begin{figure}[H]
		\centering
		\subfigure[Age groups.] 
		{
			\label{fig:pie_ageGroups}
			\includegraphics[trim =4mm 10mm 10mm 4mm, clip, width=0.25\columnwidth]{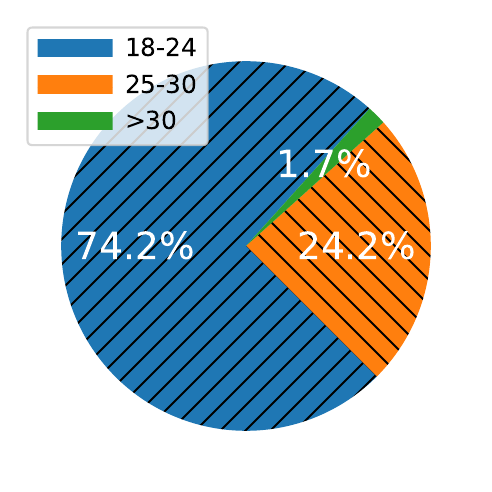}
		}
		\subfigure[Sex categories.] 
		{
			\label{fig:pie_Gender}
			\includegraphics[trim =4mm 10mm 10mm 4mm, clip, width=0.25\columnwidth]{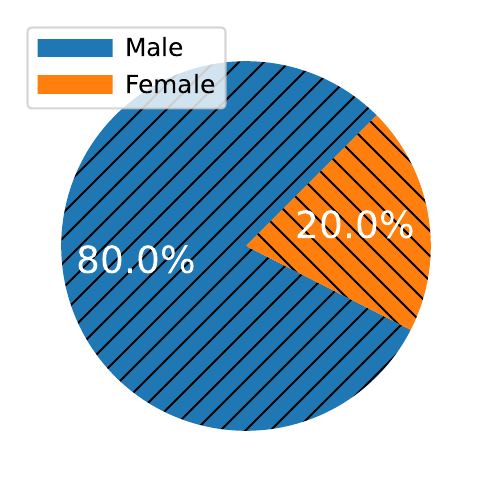}
		}
		\subfigure[Height classes.] 
		{
			\label{fig:pie_Height}
			\includegraphics[trim =4mm 10mm 10mm 4mm, clip, width=0.25\columnwidth]{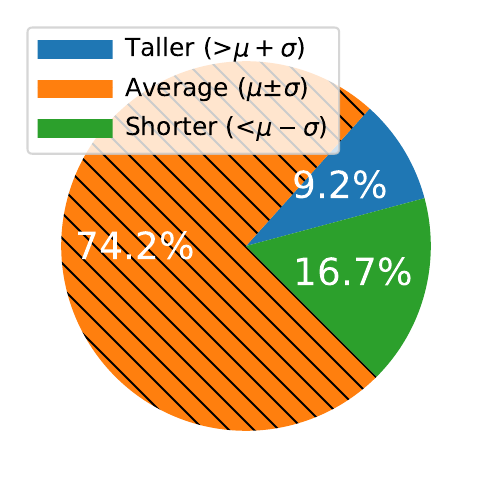}
		}
		\caption{The distribution of age, sex, and height of the volunteers.}
		\label{fig:dataDistribution}
	\end{figure}

	To collect ALS data for our experiments, we invited student volunteers to participate in our study. A total of 120~students volunteered for our study over a period of 90~days. Since the volunteers belong the student body of a large university, the majority of them were naturally in the 18-24~age group. \figurename~\ref{fig:dataDistribution} shows the distribution of the self-declared age groups, sex categories, and height classes\footnote{The volunteers declared their height classes based on the distribution of adult human heights~\cite{humanheight}.} of the volunteers.	
	\begin{figure*}[t]
		\centering
		\subfigure[From \textit{P1}.] 
		{
			\label{fig:sampleP1}
			\includegraphics[trim =0mm 0mm 0mm 0mm, clip, width=0.4\columnwidth]{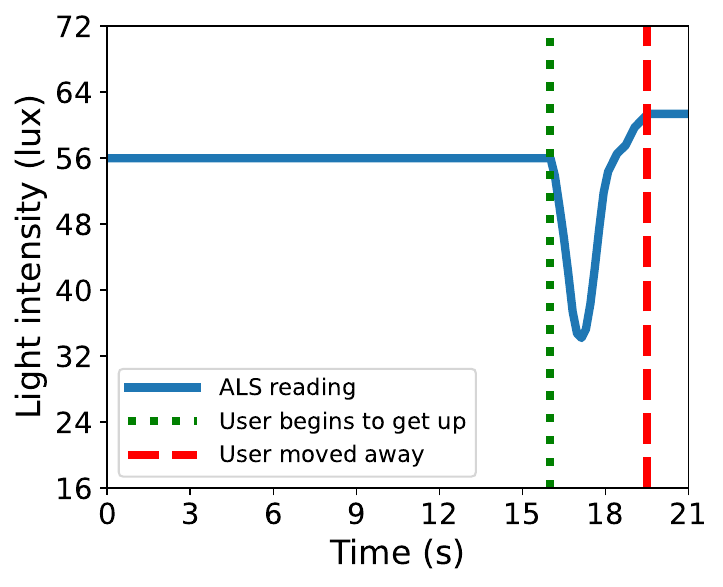}
		}
		\subfigure[From \textit{P2}.] 
		{
			\label{fig:sampleP2}
			\includegraphics[trim =0mm 0mm 0mm 0mm, clip, width=0.4\columnwidth]{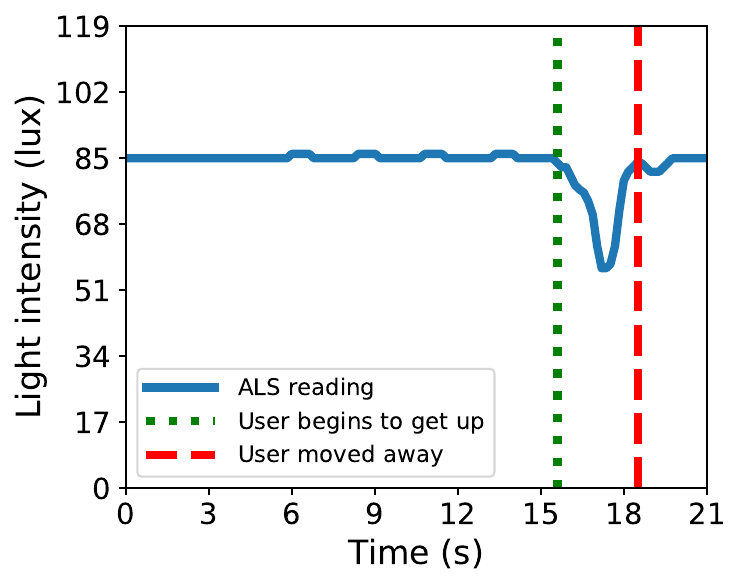}
		}
		\subfigure[From \textit{P3}.] 
		{
			\label{fig:sampleP3}
			\includegraphics[trim =0mm 0mm 0mm 0mm, clip, width=0.4\columnwidth]{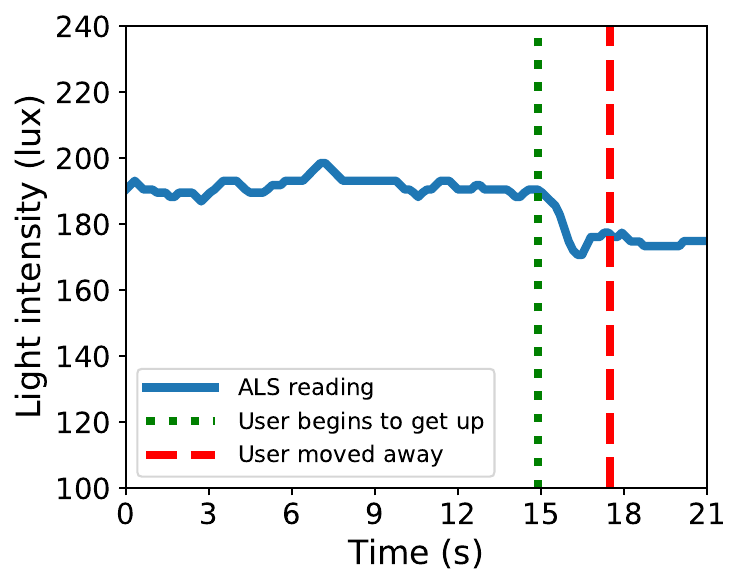}
		}
		\subfigure[From \textit{P4}.] 
		{
			\label{fig:sampleP4}
			\includegraphics[trim =0mm 0mm 0mm 0mm, clip, width=0.4\columnwidth]{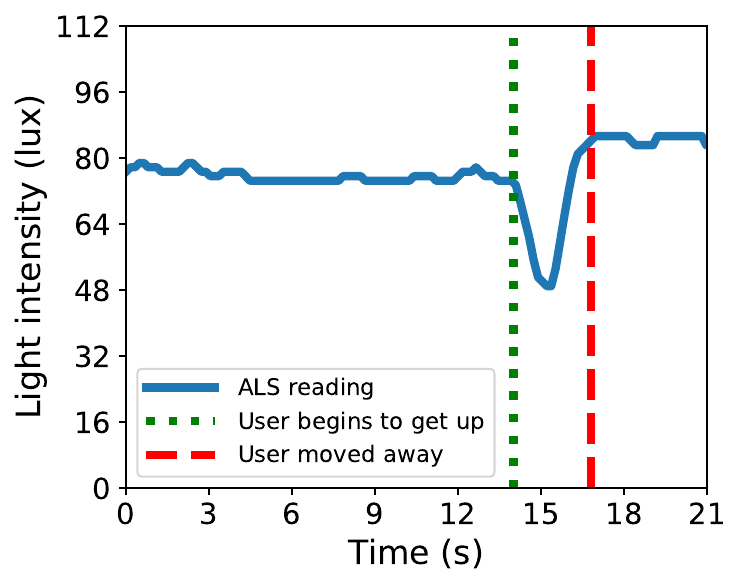}
		}
		\caption{A random set of ALS data samples collected from different positions.}
		\label{fig:sampleData}
	\end{figure*}
	\par
	Before beginning each instance of our data collection activity, we asked the volunteer to settle in a comfortable sitting/working posture at a designated desk. After which we started recording the data from the computer's ALS. At the same time, we asked the volunteer to use the computer normally for about a minute. Next, the volunteer was asked to get up and move away from the chair. It is important to highlight that to prevent any interference due to our operational activities, we remotely operated our computer to capture the data from its ALS. We also documented the time when the volunteer was instructed to get up in the activity; mainly for post-processing and analysis. Each volunteer repeated the entire activity ten times each on the four work-desk positions~(i.e., \textit{P1}, \textit{P2}, \textit{P3}, and \textit{P4}). Therefore, our dataset contains a total of 4800~data samples, i.e., 1200~data samples for each position. \figurename~\ref{fig:sampleData} depicts a random set of data samples collected from different positions during our data collection~activity. As discussed in Section~\ref{section:evalSetup}, the four work-desks experience different lighting conditions. This phenomenon is also evident from the light intensity scales shown in \figurename~\ref{fig:sampleData}~(a)-(d). We now briefly describe each plot shown in \figurename~\ref{fig:sampleData}.
	\par
	As depicted in \figurename~\ref{fig:sampleP1}, the ALS readings remain nearly constant as  long as the user remains seated in \textit{P1}. It is so because the body of the user is blocking the illumination coming from the distant light sources. The illumination perceived by ALS further drops when the user gets up. Intuitively, ALS receives a much higher illumination when the user completely moves away from the computer. 
	\par
	In case of \textit{P2} and \textit{P4}, the light sources are located on the right side and left side of the computer, respectively. ALS on our computer is located towards top-right of the screen. Thus, the chances of a sitting user shadowing LoS between ALS and the light sources are lesser in \textit{P2} when compared with \textit{P4}. As illustrated in \figurename~\ref{fig:sampleP2} and \figurename~\ref{fig:sampleP4}, both \textit{P2} and \textit{P4} observe similar levels of light intensity. However, ALS readings in \textit{P2} before and after the user moves away are at the same level, which implies that in this particular case, the user was not shadowing ALS. On another side, ALS readings in \textit{P4} after the user moves away achieve similar levels as in the case of \textit{P2}, which implies that in this particular case, the user was marginally shadowing ALS.
	\par
	Unsurprisingly, the light intensity levels are the highest in \textit{P3}. When the user moves away from \textit{P3} in the particular case shown in \figurename~\ref{fig:sampleP3}, ALS readings drop even lower than the levels when the user was sitting. A possible explanation of such a case is that the light coming from the sources in front of the user was being reflected by the user towards ALS when the user was sitting. Nevertheless, the ALS readings still clearly capture the movements of the departing user.
		
	\subsection{Experimental results}
	\label{section:results}
	We empirically assess the quality of our proposed approach with real-world data. As explained in Section~\ref{section:dataCollection}, our dataset contains a total of 4800~data samples~(i.e., 1200~data samples for each position) collected from 120~volunteers. We designed a series of experiments for a thorough analysis. We begin with investigating the general performance of DEAL. Here, we vary its input parameters to find a set of suitable configurations. Next, we study the effect of different positions~(i.e., lighting conditions) considered in our work. Finally, we examine the impact of users' height. For a de-authentication system, false negatives are more severe than false positives. At the same time, true positives are also critical. Therefore, we report the hit rate\footnote{$Recall=Hit_{Rate}=\frac{TP}{TP+FN}=1-Miss_{Rate}$} for each of our experiments.
	\par
	An analysis of our data samples indicates that the volunteers took roughly two to four seconds to get up and move away from a work-desk. Thus, we set $\ell$ between 2 and 4 seconds to approximately cover the entire user movement. We observe that a portion of the ALS readings affected due to a user's movement can be treated, depending on the value of $\Delta$, as non-outlier. It is especially witnessed for the values corresponding to the start and end of the movement; such values can still be within the threshold because the \textit{window} is not updated during a wave of consecutive outliers~(cf. lines 9, 14-17 in Algorithm~\ref{lst:algo1}). Therefore, we choose $\eta$ between 1 and 2 seconds, which is about half the time the volunteers took to move. The value of $\omega$ is fixed at 3~seconds while $\Delta$ is chosen between 5 and 20 based on preliminary experiments.
	\par
	We now discuss the generic performance of DEAL. \tablename~\ref{table:resultsOverall} shows the hit rate of our system over the entire dataset of 4800~samples for different values of $\eta$, $\ell$, and $\Delta$. An increasing value of $\Delta$ corresponds to the fact that a user should affect ALS readings substantially for the system to recognize it as an outlier. Thus, DEAL becomes resistive with increasing values of $\Delta$. Such behavior is evident in each row\footnote{E.g., $60.42 > 59.98 > 48.52 > 38.69;~\eta=1s, \ell=2s$.} of \tablename~\ref{table:resultsOverall}. The performance of our systems is affected by $\eta$ in a similar way. A larger value of $\eta$ requires a longer duration of outliers, which becomes even more challenging to attain under our stringent requirement of outliers' consecutiveness. A comparison of values\footnote{E.g., $60.42 > 50.75 > 35.10 ;~\ell=2s, \Delta=5$.} corresponding to increasing $\eta$ over fixed $\ell$ and $\Delta$ reflects the same. On another side, a larger value of $\ell$ enables us to process more values of $R$. The performance of DEAL improves with increasing value\footnote{E.g., $60.42 < 71.92 < 89.77;~\eta=1s, \Delta=5$.} of $\ell$ over a given pair of $\eta$ and $\Delta$. From these experiments, we find $\Delta=5$ and $\ell=4s$ are suitable parameter values for DEAL. Since a larger $\eta$ helps us avoid subtle user movements, we prefer $\eta=1.5s$ over $\eta=1.0s$ for our chosen values of $\Delta$ and $\ell$. With these values of  of $\Delta, \ell, \eta$, we observed a fall-out\footnote{$Fall_{Out}=FalsePositive_{Rate}=\frac{FP}{FP+TN}$} of only 7.35\%.
	\begin{table}[!htbp]
		\centering
		\caption{Hit rate~(\%) for different values of $\eta, \ell, \Delta$.}
		\label{table:resultsOverall}
		\resizebox{.75\columnwidth}{!}{
			\begin{tabular}{|c|c|c|c|c|c|}
				\hline
				\textbf{\begin{tabular}[c]{@{}c@{}}$\eta$\\ (s)\end{tabular}} & \textbf{\begin{tabular}[c]{@{}c@{}}$\ell$\\ (s)\end{tabular}} & \textbf{$\Delta=5$} & \textbf{$\Delta=10$} & \textbf{$\Delta=15$} & \textbf{$\Delta=20$} \\ \hline
				\multirow{3}{*}{1.0}                                
				& 2          & 60.42           & 59.98           & 48.52           & 38.69             \\ \cline{2-6}
				& 3          & 71.92           & 70.13           & 63.54           & 51.83             \\ \cline{2-6}
				& 4          & 89.77           & 83.67           & 69.23           & 56.50            \\ \hline
				\multirow{3}{*}{1.5}                                          
				& 2          & 50.75           & 43.79           & 32.17           & 23.69             \\ \cline{2-6}
				& 3          & 71.77           & 66.67           & 52.35           & 40.60             \\ \cline{2-6}
				& 4          & 89.15           & 74.63           & 58.42           & 45.79            \\ \hline
				\multirow{3}{*}{2.0}                                          
				& 2          & 35.10           & 24.40           & 15.90           & 10.77             \\ \cline{2-6}
				& 3          & 66.27           & 53.21           & 37.44           & 27.10             \\ \cline{2-6}
				& 4          & 86.17           & 64.04           & 45.40           & 33.65            \\ \hline
			\end{tabular}
		}
	\end{table}
	\par
	To understand the effect of different lighting conditions, we organize our dataset according to different positions~(i.e., 1200~samples per position) considered in our study. \tablename~\ref{table:resultsPosition} shows the hit rate of our system over different positions for $\eta=1.5s$, $\ell=4s$. Our results indicate that DEAL performs better in \textit{P1}, where most volunteers blocked the illumination observed by ALS while working at the computer. We see the steepest decline in the hit rate at \textit{P3}. Since \textit{P3} has copious lighting, affecting ALS readings substantially for higher values of $\Delta$ is complex. \textit{P2} and \textit{P4}, which represent normally illuminated positions and have similar light intensity levels, obtain comparable results. Overall, our system performs competently for $\Delta=5$ across different positions.
	\begin{table}[!htbp]
		\centering
		\caption{Hit rate~(\%) over different positions for $\eta=1.5s$, $\ell=4s$.}
		\label{table:resultsPosition}
		\resizebox{.75\columnwidth}{!}{
			\begin{tabular}{|c|c|c|c|c|}
				\hline
				\textbf{Position} & \textbf{$\Delta=5$} & \textbf{$\Delta=10$} & \textbf{$\Delta=15$} & \textbf{$\Delta=20$} \\ \hline
				\textit{P1}        & 91.75	    	& 82.50	    	& 67.67	    	& 56.75            \\ \hline
				\textit{P2}        & 87.00	    	& 72.08	    	& 56.08	    	& 44.75            \\ \hline
				\textit{P3}        & 90.08	    	& 70.83	    	& 53.33	    	& 35.42            \\ \hline
				\textit{P4}        & 87.75	    	& 73.08	    	& 56.58	    	& 46.25            \\ \hline
			\end{tabular}
		}
	\end{table}
	\par
\begin{figure}[H]
		\centering
		\includegraphics[trim =0mm 2mm 0mm 0mm, clip, width=0.6\columnwidth]{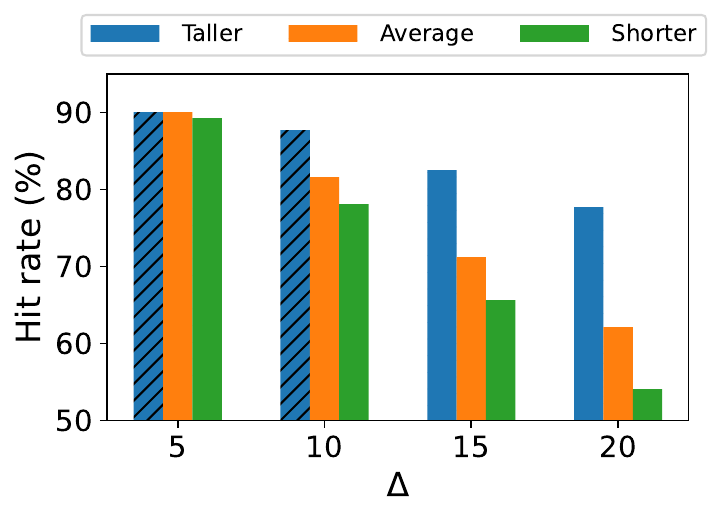}
		\caption{Hit rate~(\%) over different height classes for $\eta=1.5s$, $\ell=4s$.}
		\label{fig:resultHeight}
	\end{figure}

	Next, we consider the users' height in our study. Due to a disparity in the number of volunteers per height class, we take 250~(roughly half of the taller class samples) randomly chosen samples from each height class. \figurename~\ref{fig:resultHeight} depicts the hit rate of DEAL over different height classes for $\eta=1.5s$, $\ell=4s$. While our results for $\Delta=5$ are alike across different height classes, DEAL favors taller users for increasing values of $\Delta$. The rationale for such behavior is related to the fact that a taller user likely remains in the LoS path of ALS while working, and when such a user moves away, the ALS readings are affected sufficiently for DEAL to operate properly.
	\par
	To conclude, we argue that DEAL yields an overall effective performance. In particular, our system attains such scores without any extraordinary requirements or customization, as seen in the case of state-of-the-art solutions~(cf.~Section~\ref{section:introduction}). 
	DEAL can de-authenticate the user within two to (more realistic) four seconds~(i.e., based on the value of $\ell$). 
	In a real-world deployment, an enrollment step at the end user's work-desk can help tune the system to function even better.

	\section{Discussion}
	\label{section:discussion}
	We specify the key attributes of our work in this section. Section~\ref{section:comparison} compares DEAL with state-of-the-art user de-authentication schemes and highlights its salient features. Section~\ref{section:limitations} discusses the potential limitations of DEAL.
	\subsection{Comparison with existing de-authentication schemes}
	\label{section:comparison}
	For a rigorous comparison among the key de-authentication solutions, we assess each one of them on a dozen crucial dimensions. \tablename~\ref{table:comparison} summarizes our comparison and underlines the prominent features and limitations of the key existing solutions.
	
	\begin{table*}[!htbp]
		\centering
		\caption{A comparative summary of the key de-authentication schemes.}
		\label{table:comparison}
		\resizebox{1.99\columnwidth}{!}{
			\begin{tabular}{|l|c|c|c|c|c|c|c|c|c|c|c|l|r|}
				\cline{1-14}
				\multicolumn{1}{|c|}{\circled{\textbf{\scriptsize 0}}}                                                   & \circled{\textbf{\scriptsize 1}}                                   & \circled{\textbf{\scriptsize 2}}                        & \circled{\textbf{\scriptsize 3}}                      & \circled{\textbf{\scriptsize 4}}                             & \circled{\textbf{\scriptsize 5}}                       & \circled{\textbf{\scriptsize 6}}                             & \circled{\textbf{\scriptsize 7}}                           & \circled{\textbf{\scriptsize 8}}                                     & \circled{\textbf{\scriptsize 9}}                         & \circled{\textbf{\scriptsize 10}}                                                 & \circled{\textbf{\scriptsize 11}}                       & \multicolumn{1}{c|}{\circled{\textbf{\scriptsize 12}}} & \multicolumn{1}{c|}{\circled{\textbf{\scriptsize 13}}} \\
				\hline	
				{\parbox{1.00cm}{\textbf{Scheme}}}  
				& \rot{\parbox{2.5cm}{\textbf{Unobtrusive}}} 
				& \rot{\parbox{2.5cm}{\textbf{Nothing to hold}}} 
				& \rot{\parbox{2.5cm}{\textbf{Non-biometric}}} 
				& \rot{\parbox{2.3cm}{\textbf{Non-continuous authentication}}} 
				& \rot{\parbox{2.5cm}{\textbf{Enrollment-free}}} 
				& \rot{\parbox{2.5cm}{\textbf{Difficult to evade}}} 
				& \rot{\parbox{2.55cm}{\textbf{Inactivity supported}}} 
				& \rot{\parbox{2.2cm}{\textbf{No external equipment needed}}} 
				& \rot{\parbox{2.5cm}{\textbf{Maintenance-free}}} 
				& \rot{\parbox{2.0cm}{\textbf{No particular installation or customization}}} 
				& \rot{\parbox{2.5cm}{\textbf{Simple to deploy}}} 
				& \rot{\parbox{2.5cm}{\textbf{Price}}} 
				&  \rot{\parbox{2.2cm}{\textbf{Subjects in user study}}} \\ \hline    		
				Timeout                                           & \cmark                                  & \cmark                       & \cmark                     & \cmark                                     & \cmark                       & \xmark                          & \xmark                            & \cmark                                    & \cmark                        & \cmark                                                   & \cmark                        & Low (free)         & --                                        \\ \hline
				ZEBRA~\cite{mare2014zebra}                        & \xmark                                  & \xmark                       & \cmark                     & \xmark                                     & \xmark                       & \xmark                          & \xmark                            & \xmark                                    & \xmark                        & \cmark                                                   & \cmark                        & Medium (\$100-200) & 20                                        \\ \hline
				Gaze tracking~\cite{eberz2015preventing}          & \cmark                                  & \cmark                       & \xmark                     & \xmark                                     & \xmark                       & \cmark                          & \xmark                            & \xmark                                    & \cmark                        & \xmark                                                   & \cmark                        & High (\$2-5k)      & 30                                        \\ \hline
				Assentication~\cite{kaczmarek2018assentication}   & \cmark                                  & \cmark                       & \xmark                     & \xmark                                     & \xmark                       & \cmark                          & \cmark                            & \xmark                                    & \xmark                        & \xmark                                                   & \xmark                        & Medium (\$150)     & 30                                        \\ \hline
				Pulse-response~\cite{rasmussen2014authentication} & \xmark                                  & \xmark                       & \xmark                     & \xmark                                     & \xmark                       & \cmark                          & \cmark                            & \xmark                                    & \cmark                        & \cmark                                                   & \xmark                        & Unknown            & 10                                        \\ \hline
				Keystroke dynamics~\cite{joyce1990identity}       & \cmark                                  & \cmark                       & \xmark                     & \xmark                                     & \xmark                       & \cmark                          & \xmark                            & \cmark                                    & \cmark                        & \cmark                                                   & \cmark                        & Low (free)         & 33                                        \\ \hline
				FADEWICH~\cite{conti2017fadewich}                 & \cmark                                  & \cmark                       & \cmark                     & \cmark                                     & \cmark                       & \xmark                          & \cmark                            & \xmark                                    & \cmark                        & \xmark                                                   & \xmark                        & Unknown            & 3                                         \\ \hline
				DEB~\cite{conti2020auth}                          & \cmark                                  & \cmark                       & \cmark                     & \cmark                                     & \cmark                       & \cmark                          & \cmark                            & \xmark                                    & \xmark                        & \xmark                                                   & \cmark                        & Low (\$10)         & 15                                        \\ \hline
				BLUFADE~\cite{blufade}                            & \cmark                                  & \cmark                       & \xmark                     & \xmark                                     & \xmark                       & \cmark                          & \cmark                            & \cmark                                    & \cmark                        & \xmark                                                   & \cmark                        & Low (\$5)         & 30                                        \\ \hline
				ZIA~\cite{corner2002zero}                         & \xmark                                  & \xmark                       & \cmark                     & \xmark                                     & \xmark                       & \cmark                          & \cmark                            & \xmark                                    & \xmark                        & \cmark                                                   & \cmark                        & Low (\$10-30)      & 1                                         \\ \hline
				1DMRLBP~\cite{ecg}                                & \xmark                                  & \xmark                       & \xmark                     & \xmark                                     & \xmark                       & \cmark                          & \cmark                            & \xmark                                    & \cmark                        & \cmark                                                   & \cmark                        & Medium (\$50-200)  & --                                        \\ \hline
				\cellcolor{cellFillingColor}DEAL~(our proposal)                                             & \cellcolor{cellFillingColor}\cmark                       & \cellcolor{cellFillingColor}\cmark                     & \cellcolor{cellFillingColor}\cmark                            & \cellcolor{cellFillingColor}\cmark                            & \cellcolor{cellFillingColor}\cmark                                  & \cellcolor{cellFillingColor}\cmark                          & \cellcolor{cellFillingColor}\cmark                       & \cellcolor{cellFillingColor}\cmark                                    & \cellcolor{cellFillingColor}\cmark                        & \cellcolor{cellFillingColor}\cmark                                                  & \cellcolor{cellFillingColor}\cmark                        & \cellcolor{cellFillingColor}Low (free)         & \cellcolor{cellFillingColor}120 \\ \hline
			\end{tabular}
		}
	\end{table*}
	\par
	One of the fundamental requirements for any consumer technology is its user-friendliness. In our context, it is directly related to the unobtrusiveness~(cf.~col.~\circled{\textbf{\scriptsize 1}})  of a given de-authentication solution and whether it compels the user to carry, wear, or bear anything extra~(cf.~col.~\circled{\textbf{\scriptsize 2}}). We find that ZEBRA, pulse-response, ZIA, and 1DMRLBP can cause inconvenience to the user by requiring them to bear a bracelet, a pair of electrodes, a token, and an ECG apparatus, respectively. The existing solutions can be further classified as biometric or non-biometric~(cf.~col.~\circled{\textbf{\scriptsize 3}}) and continuous\footnote{The user is re-authenticated throughout the session, and de-authentication happens whenever she cannot prove her identity.} or non-continuous solutions~(cf.~col.~\circled{\textbf{\scriptsize 4}}). Biometric-based solutions~(i.e., gaze tracking, \textit{Assentication}, pulse-response, key-stroke dynamics, BLUFADE, 1DMRLBP) are certainly difficult to evade~(cf. col.~\circled{\textbf{\scriptsize 6}}) as imitating someone else's biometry or behavioral patterns is highly complex. On the other side, some continuous solutions can be subverted. For instance, authors in the work~\cite{huhta2015pitfalls} have shown that an attacker can evade ZEBRA via opportunistic observations. Both the biometric and continuous solutions are accurate. However, the performance of both the categories of solutions comes at the cost of:~(1)~a user enrollment phase~(cf.~col.~\circled{\textbf{\scriptsize 5}}) that can be laborious and time-consuming for the end-user, and (2)~the cost of equipment required to capture their respective features is non-trivial. Only a few solutions are enrollment-free. Regarding the difficulty of evasion, FADEWICH is not suitable for a densely occupied workspace, while the classic timeout approach fails to sense the user's absence.
	\par
	A user may not interact continuously with her computer~(e.g., while attending a phone call). Thus, another key attribute of a user-centric de-authentication scheme is its support for a user's inactivity~(cf.~col.~\circled{\textbf{\scriptsize 7}}). Timeout, ZEBRA, gaze tracking, and keystroke dynamics depend on user interactions, and thus they violate this objective. The main limitation of the majority of existing schemes is their dependence on external equipment for operation~(cf.~col.~\circled{\textbf{\scriptsize 8}}). Such a dependence not only hinders their widespread adoption, but it can also spawn several related concerns, i.e., maintenance, physical customization, deployment complexity, and price. Only timeout approach, keystroke dynamics, BLUFADE, and our proposal do not depend on external hardware; thus, they are not generally affected by the consequent concerns mentioned~before. 
	\par
	ZEBRA, DEB, and ZIA demand periodic recharging or replacement of batteries while \textit{Assentication} requires maintenance of the wires that supply power to the chair. The external hardware in the other such solutions is powered directly by the target computer. Finally, all these solutions also involve the risk of physical damage to the external hardware that may seek a replacement~(cf.~col.~\circled{\textbf{\scriptsize 9}}).
	\par
	Some of the solutions that use external hardware require a particular installation of the equipment~(i.e., gaze tracking, DEB) or even customization to workplace infrastructure~(i.e., \textit{Assentication}, FADEWICH). The user simply holds/wears the external apparatuses in other such solutions~(i.e., ZEBRA, ZIA, 1DMRLBP, pulse-response). As discussed in Section~\ref{section:relatedWork}, BLUFADE requires affixing a particular physical barrier on the webcam~(cf.~col.~\circled{\textbf{\scriptsize 10}}). Regarding deployment complexity~(cf.~col.~\circled{\textbf{\scriptsize 11}}), \textit{Assentication} and FADEWICH are not simple to deploy in practice as they require alteration to infrastructure. Similarly, pulse-response involves complex handling of multiple apparatuses~(arbitrary waveform generator, oscilloscope, brass hand-electrode, etc.). All the remaining solutions are simple to deploy even when they need particular placement of hardware~(e.g., gaze tracking, DEB, BLUFADE). As far as the price is concerned~(cf.~col.~\circled{\textbf{\scriptsize 12}}), gaze tracking employs an expensive eye-tracking device. Though the price of FADEWICH and pulse-response is unknown, we suppose they are slightly costlier as they use several sensors and apparatuses. The cost of the remaining schemes is low to medium. Finally, the number of subjects in the user/validation study could indicate the robustness of evaluation results, which is the highest in our case~(cf.~col.~\circled{\textbf{\scriptsize 13}}).
	\par
	In light of our analysis, we find that the state-of-the-art solutions lack a few or several vital characteristics of an effective and practical de-authentication scheme. On the other hand, DEAL is the only solution that possesses all these characteristics. Therefore, we believe it is the most useful and practical de-authentication scheme.
	
	\subsection{Limitations}
	\label{section:limitations}
	We now ponder upon the potential limitations of DEAL.
	\begin{enumerate}
		\item \textit{ALS' presence:}
		Our proposed de-authentication approach relies on an ALS. While ALS has been present on smartphones and tablets for a long time, it has only recently started to become available on laptops~(e.g., MacBooks) and desktops~(e.g., iMacs). Therefore, DEAL is futuristic and suitable for newer generations of computers. Nevertheless, one can attach a USB-powered ALS to use DEAL in the absence of a built-in ALS.
		\par
		One related issue could be the physical placement of ALS on the computer. ALS~(like other user-centric sensors, e.g., webcam) is generally mounted on the front side of the display screen. Our approach will work as long as ALS faces the users. For the sake of readers' convenience, \figurename~\ref{fig:concept} and \figurename~\ref{fig:conceptDesktop} conceptualize DEAL on a laptop and a desktop, respectively.
		\begin{figure}[H]
			\centering
			\includegraphics[trim =2mm 4mm 3mm 0mm, clip, width=0.32\columnwidth]{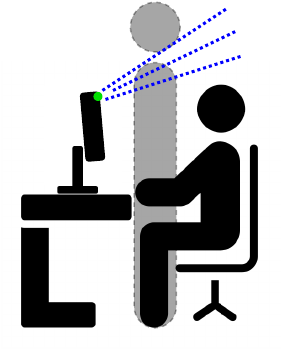}
			\caption{A representative conceptualization of DEAL on a desktop.}
			\label{fig:conceptDesktop}
		\end{figure}
		However, any unusual sensor placement~(e.g., behind the screen panel) will render our system unusable. It is worth noting that such an unusual sensor placement could be suitable for portable devices, but not for computers that can be docked near a wall.
		\item \textit{False alarms due to passersby:}
		A common phenomenon in any workplace setting is the movements of passersby~(e.g., colleagues). We set up a separate experiment to investigate such a scenario. \figurename~\ref{fig:NearBy} shows different user positions, where \circled{\textbf{\scriptsize A}} represents the legitimate user's standing position, \circled{\textbf{\scriptsize B}} shows a passerby crossing too close to the target user, and \circled{\textbf{\scriptsize C}} depicts a passerby away from the target user.
		\begin{figure}[H]
			\centering
			\includegraphics[trim =3mm 4mm 0mm 0mm, clip, width=0.5\columnwidth]{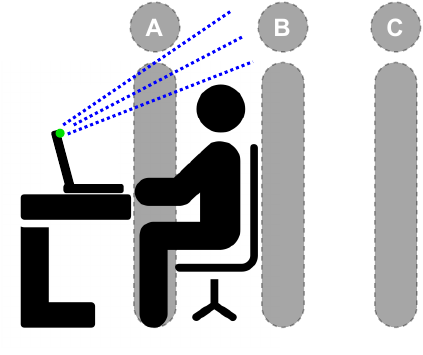}
			\caption{An illustration of passersby near our user's work-desk.}
			\label{fig:NearBy}
		\end{figure}
		\par
		We find that our system remains largely unaffected as long as a passerby~(cf. \circled{\textbf{\scriptsize C}}) walks at some~(about 2-3~ft) distance from the user. In particular, any wave of outliers, if induced, is sparse and short. On the other hand, our system de-authenticates the user when a passerby~(cf.~\circled{\textbf{\scriptsize B}}) comes too close to the user; as it affects the ALS readings. It can be seen as a false alarm. Nevertheless, such de-authentications can protect the user's privacy from shoulder surfers and onlookers.
		
		\item \textit{Violation of our assumptions:}
		Our system will create a false alarm if the user changes her working posture~(e.g., from sitting to standing). Similarly, it may possibly not de-authenticate the user if she moves away from her work-desk without getting up~(e.g., by dragging the chair). Violating the assumptions or requirements of any given scheme will affect its functioning, and our work is no different. Nevertheless, one may see it as a limitation of our work.
	\end{enumerate}
	
	\section{Conclusion and future work}
	\label{section:conclusionFutureWork}
	Both user authentication and de-authentication are essential operations for the security of a computer system. It is even more critical to de-authenticate a user in a shared workspace setting because an insider can gain access to the user's active session through \textit{lunchtime} attacks. The research community has proposed different de-authentication and continuous-authentication techniques over the inactivity timeout-based method. The existing works unfortunately have various limitations, e.g., complex installation procedures, requirement of external hardware to assert user presence. In this paper, we propose a novel approach, called DEAL, that uses ALS present on a computer to de-authenticate the user. We assessed the quality of our proposed approach empirically in the real world. While being effective and fast, DEAL is also unobtrusive.
	\par
	In the future, we would like to test DEAL in unconventional workplace settings, such as in a cafe or under different colored lighting. We will explore the possibility of assisting DEAL with machine learning-based classification techniques to further improve its performance. We will also investigate the effect of personalized tuning~(e.g., via an enrollment stage for the end user) on its performance.
	
	\section*{IRB approval}
	We obtained prior approval for our experiments from the Institutional Review Board~(IRB) of the institute, where the experiments were carried out. The level of review recommendation was: \textit{Exempt.} All participants were volunteers, who were informed of the actual use of the collected data, and their informed consent was obtained before starting the data collection process. No sensitive data was collected. In particular, no participant names, contact numbers, or other Personally Identifying Information~(PII) was collected. The minimal identifying information retained was also anonymized. All the data was~(and is) stored in an encrypted form.
	\bibliographystyle{IEEEtran}
	\balance
	\bibliography{bib}
\end{document}